\documentclass[floatfix,reprint,twocolumn,superscriptaddress,showpacs,nofootinbib,notitlepage,aps,prl,preprintnumbers]{revtex4-1}
\usepackage{pgfplots}
\pgfplotsset{width=3cm,compat=newest}

\usepackage[utf8]{inputenc}
\usepackage{graphicx}
\usepackage{latexsym,amsmath,amssymb,lmodern,url}
\usepackage{natbib}
\usepackage{color}
\usepackage{microtype}
\usepackage{slashed}
\usepackage{multirow}
\usepackage{bbm}
\usepackage{xcolor}
\usepackage{tikz}
\usetikzlibrary{shapes}

\usepackage[colorlinks=true,backref=false, linktocpage=true,
citecolor=blue,urlcolor=blue,linkcolor=blue,pdfpagemode=UseOutlines]{hyperref}

\hypersetup{%
  bookmarksnumbered=true,
  pdftitle = {},
  pdfsubject = {},
  pdfauthor = {},
  pdfkeywords = {}
}

\let\Re\undefined

\newcommand{\ds}{S({1080})}
\DeclareMathOperator{\Tr}{Tr}
\DeclareMathOperator{\Re}{Re}

\newcommand{\id}{\mathbbm 1}

\definecolor{mycolor}{RGB}{22,139,22}
\newcommand{\bluefcir}{\raisebox{0.7pt}{\tikz{\node[draw,scale=0.4,circle,draw=blue,fill=blue,rotate=0](){};}}}
\newcommand{\redfcir}{\raisebox{0.7pt}{\tikz{\node[draw,scale=0.4,circle,draw=red,fill=red,rotate=0](){};}}}
\newcommand{\blackcir}{\raisebox{0.7pt}{\tikz{\node[draw,scale=0.4,circle,draw=black,fill=black!20!black,rotate=0](){};}}}
\newcommand{\blackecir}{\raisebox{0.7pt}{\tikz{\node[draw,scale=0.4,circle,draw=black,fill=white,rotate=0](){};}}}

\begin{document}
\preprint{FERMILAB-PUB-20-045-T}
\title{Quantum Simulation of Field Theories Without State Preparation}
\author{Siddhartha Harmalkar}
\email{sharmalk@umd.edu}
\affiliation{Department of Physics, University of Maryland, College Park, MD 20742, USA}
\author{Henry Lamm}
\email{hlamm@fnal.gov}
\affiliation{Fermi National Accelerator Laboratory, Batavia,  Illinois, 60510, USA}
\author{Scott Lawrence}
\email{srl@umd.edu}
\affiliation{Department of Physics, University of Maryland, College Park, MD 20742, USA}
\date{\today}
\collaboration{NuQS Collaboration}\noaffiliation
\begin{abstract}
We propose an algorithm for computing real-time observables using a quantum processor while avoiding the need to prepare the full quantum state. This reduction in quantum resources is achieved by classically sampling configurations in imaginary-time using standard lattice field theory.  These configurations can then be passed to quantum processor for time-evolution. This method encounters a signal-to-noise problem which we characterize, and we demonstrate the application of standard lattice QCD methods to mitigate it.
\end{abstract}

\maketitle


Nonperturbative computational methods for quantum field theories are limited by an inability to access real-time observables, such as conductivities and viscosities. Stochastic methods (in particular lattice QCD) suffer from a maximally bad sign problem \cite{Alexandru:2016gsd}, whereas deterministic methods are infeasible due to the exponential growth of the Hilbert space with volume. The creation of quantum processors promises access to these nonperturbative observables: quantum processors, unlike their classical counterparts, can efficiently represent and manipulate the exponentially large Hilbert space.

A natural way to simulate a quantum mechanical system on a quantum processor is to map the physical Hilbert space to that of the processor, and then perform unitary operations that mimic time-evolution~\cite{Lloyd1073}. This strategy works for lattice-regularized quantum field theories~\cite{Ortiz:2000gc,Zohar:2016iic,Lamm:2019bik,Klco:2019evd}. The primary theoretical difficulty for such simulations is efficiently creating the initial, strongly-coupled quantum state on the quantum processor.  Current algorithms for this state preparation step are expensive to implement (usually dominating the circuit depth of simulations), difficult to analyze, and often unlikely to generalize.  In particular, preparing scattering states requires substantial complexity~\cite{Jordan:2011ne,Jordan:2011ci,Garcia-Alvarez:2014uda,Jordan:2014tma,Jordan:2017lea,Moosavian:2017tkv,Moosavian:2019rxg,Gustafson:2019vsd}.  Consequently, many works have studied the problem, of which~\cite{Kokail:2018eiw,Lamm:2018siq,Klco:2019xro,Klco:2019yrb} is a small sample. Here, we show how to offload this step to a classical computer, reducing the quantum circuit depth.

In~\cite{Lamm:2018siq}, two of us argued that preparing the full quantum state on a quantum processor was unnecessary.  Instead, the quantum processor could be used as a black-box oracle for calculating real-time matrix elements between basis states which are sampled based on a classical Monte Carlo calculation of the density matrix.  However, that method relied on Density Matrix Quantum Monte Carlo (DMQMC)~\cite{PhysRevB.89.245124} which is unwieldy for lattice field theory.

In this Letter, we present a simpler algorithm using standard tools from lattice field theory for the classical portion of the computation. The algorithm proceeds as a Monte Carlo calculation in Euclidean lattice field theory, invoking a quantum processor only to compute matrix elements of time-dependent operators. Our method suffers from a classical signal to noise (StN) problem of a more general kind than the usual Parisi-Lepage type~\cite{Parisi:1983ae,Lepage:1989hd}. Existing lattice field theory techniques can address this StN problem. We discuss some of these modifications and their incorporation into our algorithm if necessary for evaluating real-time correlators on near-term processors.

Consider a thermal state given by a density matrix $\rho\equiv e^{-\beta H_0}$ with a Hamiltonian $H_0$ and inverse temperature $\beta$.  Of interest is the response of this state to time-evolution by $e^{iH_1 t}$ ($H_1$ need not be close to $H_0$).   
The expectation value of $\mathcal O(t)\equiv e^{i H_1 t} \mathcal O e^{-i H_1 t}$ is given by:
\iftrue
\begin{widetext}
\begin{equation}\label{eq:expectation}
\left<\mathcal O(t)\right> = \frac{\Tr e^{-\beta H_0} \mathcal O(t)}{\Tr e^{-\beta H_0}}
=\frac
{\sum_{i,j}
\rho_{ji}
\mathcal O(t)_{ij}
}{\sum_{i}\rho_{ii}}
=\left(
\frac
{\sum_{i,j}
\rho_{ji}
\mathcal O(t)_{ij}
}
{\sum_{i,j}\rho_{ij}}
\right)
\left(
\frac
{\sum_i \rho_{ii}}
{\sum_{i,j} \rho_{ij}}
\right)^{-1}
\equiv\frac{\langle \mathcal O(t)\rangle_{\rho}}{\langle \delta_{ij}\rangle_{\rho}}
\end{equation}
\end{widetext}
\else
\begin{align}\label{eq:expectation}
&\left<\mathcal O(t)\right> = \frac{\Tr e^{-\beta H_0} e^{i H_1 t} \mathcal O e^{-i H_1 t}}{\Tr e^{-\beta H_0}}\nonumber\\
&\hspace{1cm}=\frac
{\sum_{i,j}
\left<\Psi_j\right|\rho\left|\Psi_i\right>
\left<\Psi_i\right|\mathcal O(t)\left|\Psi_j\right>
}{\sum_{i}\left<\Psi_i\right|\rho\left|\Psi_i\right>}\nonumber\\
=&\bigg(
\frac
{\sum_{i,j}
\langle\Psi_j|\rho|\Psi_i\rangle
\langle\Psi_i|\mathcal O(t)|\Psi_j\rangle
}
{\sum_{i,j}\langle\Psi_j|\rho|\Psi_i\rangle}
\bigg)
\left(
\frac
{\sum_i \langle\Psi_i|\rho|\Psi_i\rangle}
{\sum_{i,j}\langle\Psi_j|\rho|\Psi_i\rangle}
\right)^{-1}\nonumber\\
&\hspace{2.5cm}\equiv\frac{\langle \mathcal O(t)\rangle_{\rho}}{\langle \delta_{ij}\rangle_{\rho}}
\end{align}
\fi
where $\rho_{ij}$ and $\mathcal O(t)_{ij}$ respectively denote matrix elements $\left<\Psi_i\right|\rho\left|\Psi_j\right>$ and $\left<\Psi_i\right|\mathcal O(t)\left|\Psi_j\right>$, with the $\Psi_i$ states in a basis easily prepared on the quantum processor.  Critically, the basis states are cheap to prepare. The notation $\langle \mathcal \cdot\rangle_\rho$ denotes expectation values sampled from the distribution $\rho_{ij}\equiv\langle\Psi_j|\rho|\Psi_i\rangle$.  The overall normalization $\langle \delta_{ij}\rangle_{\rho}$ measures the weight of $\sum_i\rho_{ii}$; while often unneeded, an efficient classical procedure for it is in the appendix.

\begin{figure*}
\includegraphics[width=0.9\linewidth]{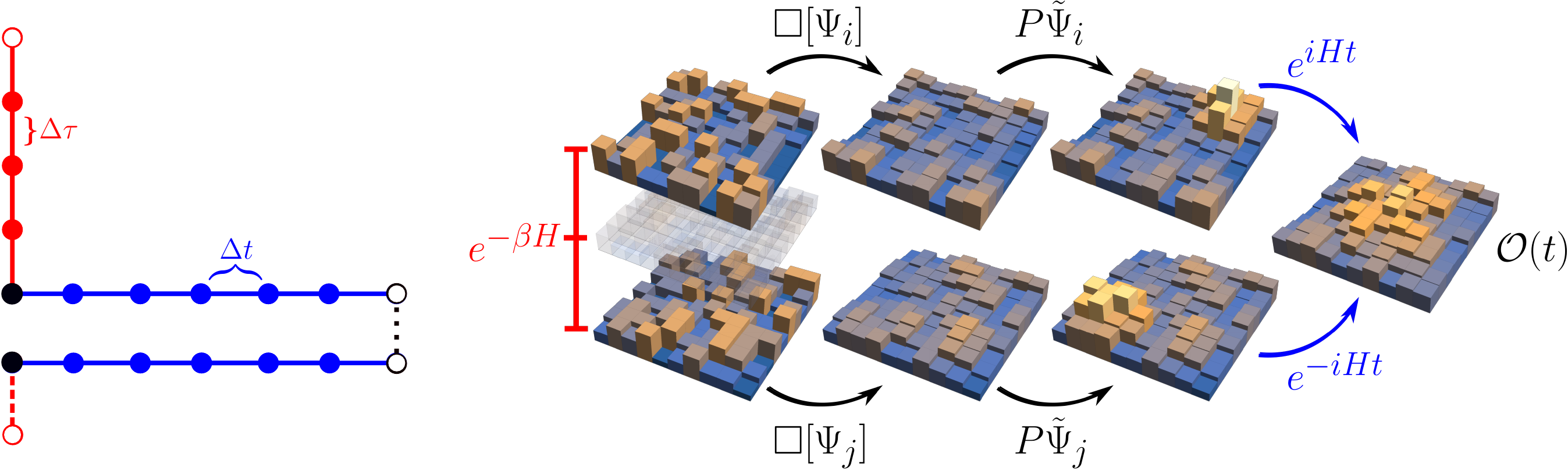}
\caption{\label{fig:sk} Two schematic views: (left) the Schwinger-Keldysh contour in the complex time plane with the $\tau$ path integral~(\protect \redfcir) discretized with lattice spacing $\Delta \tau$ and the $t$ path integral~(\protect \bluefcir) trotterized with step $\Delta t$.~(\protect \blackcir) are the matching points between the two path integrals, while~(\protect \blackecir) correspond to inserting $\mathcal{O}$ (right) cartoon of how the StN can be improved through smearing, $\Box[\Psi]=\tilde{\Psi}$, and interpolators $P$ permit the preparation of configurations overlapping with non-thermal states}
\end{figure*}

The connection between Eq.~(\ref{eq:expectation}) and lattice field theory may be seen via the Schwinger-Keldysh formalism~\cite{Schwinger:1960qe,Keldysh:1964ud}, where $\langle \mathcal O(t)\rangle$ is given as a path integral in both real time $t$ and imaginary time $\tau$:
\begin{equation}\label{eq:ski}
 \langle\mathcal{O}(t)\rangle=\frac{1}{Z}\int\mathcal D\Psi \;e^{iS_{SK}[\Psi]}\;\mathcal{O}(t)
\end{equation}
where $S_{SK}=\int\mathrm{d}t L_1[x]+\int\mathrm{d}\tau L_0[x]\equiv S_1[\Psi(t)]+S_0[\Psi(\tau)]$ is integrated along a contour $\mathcal{C}$ in both $t$ and $\tau$, shown on the left of Fig.~(\ref{fig:sk}). This closed contour can be decomposed into coupled open contours that are purely Minkowski and Euclidean, respectively\footnote{Since the path integrals have different arguments $T,H_1,\Delta t, etc.$ versus $\beta, H_0, \Delta \tau, etc.$, renormalization factors will generally differ, and must be computed separately.}. These open contours correspond to the matrix elements $\rho_{ji}$ and $\mathcal O(t)_{ij}$, and the states at the `corners' of the SK contour are $\Psi_i$ and $\Psi_j$.  To recover Eq.~(\ref{eq:ski}), one takes a sum over these states.

The matrix elements of $\mathcal O(t)$ may be efficiently computed on a quantum processor~\cite{PhysRevLett.118.010501,PhysRevLett.123.070503,Roggero:2018hrn,Zohar:2018cwb,Clemente:2020lpr}.
The elements of $\rho$ correspond to the Euclidean part of the path integral, with state $\Psi_i$ at the bottom of the temporal direction and $\Psi_j$ at the top.  Standard lattice field theory is an effective technique for calculations involving this object. For operators $\mathcal O$ diagonal in the computational basis, the the sum over $j$ in Eq.~(\ref{eq:expectation}) may be eliminated, and the expectation value $\langle\mathcal O\rangle$ is evaluated by sampling with respect to the diagonal of $\rho$. In contrast, the time-evolved $\mathcal O(t)$ is far from diagonal in our basis, and so we must retain the sums over both $\Psi_i$ and $\Psi_j$.

Thus, in Eq.~\ref{eq:expectation}, the physical expectation value $\langle\mathcal O(t)\rangle$ is written as a ratio of two expectation values taken with respect to the full density matrix (i.e., the distribution on pairs of states given by $\rho_{ij}$). When sampling with respect to the diagonal of $\rho$, one requires that the $\Psi_i=\Psi_j$ , thus imposing periodic boundary conditions (PBC) in $\tau$. When sampling from the full $\rho$ instead, both the $\Psi_i$ and $\Psi_j$ are summed over, corresponding to open boundary conditions (OBC). Between these, the operator $\mathcal O(t)$ is inserted by querying the quantum processor for $\mathcal O(t)_{ij}$.

OBCs in lattice QCD were first proposed in~\cite{Luscher:2011kk} to study topology and autocorrelation.  Other uses were spectroscopy~\cite{Luscher:2012av}, scale setting~\cite{Hollwieser:2019kuc}, and topological susceptibility~\cite{Florio:2019nte}. We advocate their use to compute $\mathcal O(t)$.

Our algorithm, then, is to perform a classical lattice Monte Carlo with OBC, coupled to a quantum processor which evaluates $\mathcal O(t)_{ij}$ on each configuration (with $\Psi_i$ and $\Psi_j$ the initial and final $\tau$-slices of the Euclidean lattice). We stress that the imaginary time evolution is performed on the classical machine, and the quantum processor only needs enough qubits available to store a single state (i.e. a single timeslice of the Euclidean lattice).

So far, the discussion has focused on thermal states.  Using Euclidean lattice techniques, we can extend to other states.  With configurations sampled from $\rho$, Euclidean observables are equal to sums over complete set of eigenstates $|n\rangle$ with weights $e^{-E_n\beta}$.  If we want the matrix elements $\langle\psi|\mathcal O|\psi\rangle$ of a particular state $|\psi\rangle$, it is necessary to isolate it from all other states. When $|\psi\rangle$ is the lowest state with the quantum numbers of some projection operator $P$, the procedure is simple: insert a source $P$ and a sink $P^\dag$ between $\rho$ and $\mathcal{O}$, and consider large $\tau$:
\begin{align}
\sum_{n}&\langle n |P^\dag \rho P |n\rangle\langle n|\mathcal{O}|n\rangle\nonumber\\
&=\langle \psi|\mathcal{O}|\psi\rangle\langle \psi |P^\dag P|\psi\rangle e^{-E_{\psi}\tau}+\mathcal{O}(e^{-\Delta E \tau}).
\end{align}
where exponentially-suppressed corrections come from other eigenstates, $|n\rangle$, that overlap with $P$.  To extract $\langle\psi\mathcal| \mathcal O|\psi\rangle$, one notes that $\langle \psi |P^\dag \rho P|\psi\rangle$ can be obtained from a two-point correlator $\langle P^\dag(\tau)P(0)\rangle$.  This method can initialize the quantum processor with configurations that overlap with the desired state, avoiding direct preparation.

\begin{figure}
\begin{center}
\includegraphics[width=\linewidth]{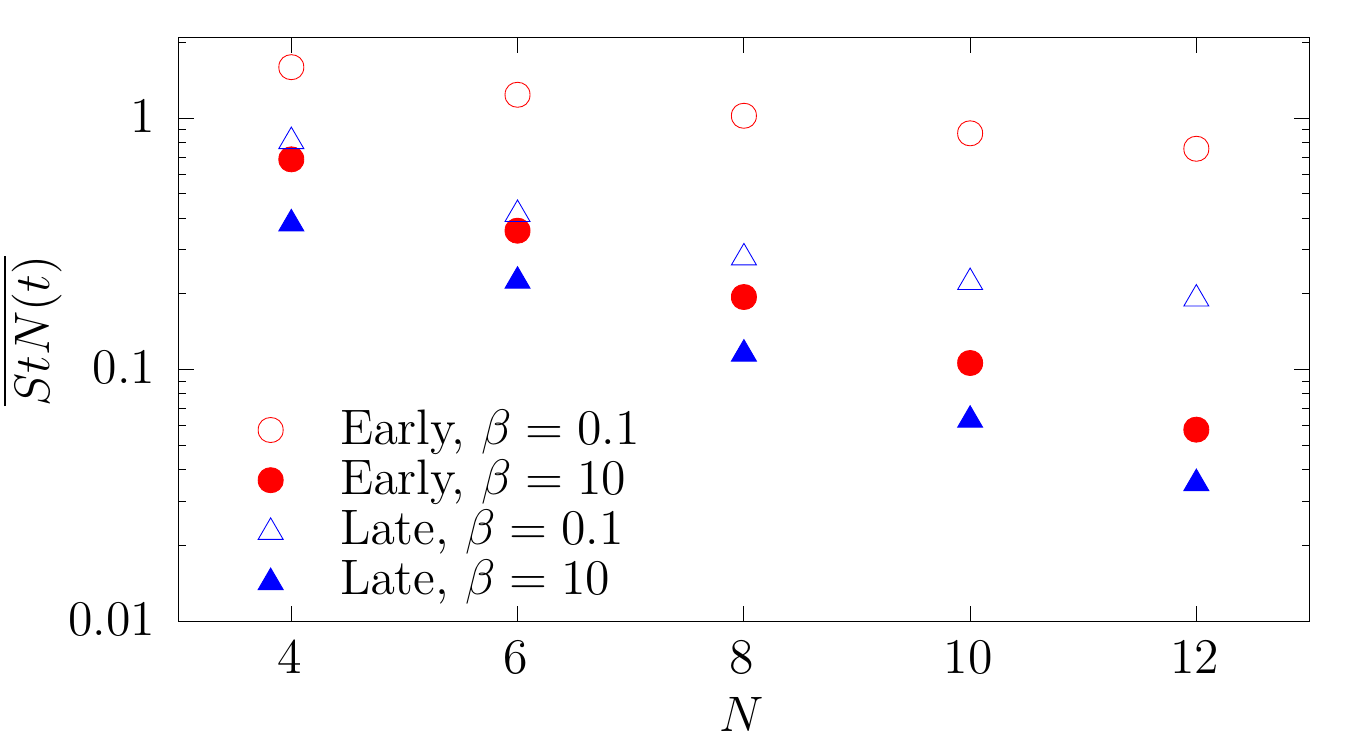}
\end{center}
\caption{\label{fig:stn}Average value of the signal-to-noise over early ($t \in [0,0.1]$) and late ($t \in [5,15]$) times, at two inverse temperatures $\beta$. The StN decreases exponentially with the size of the system, and is larger for higher temperatures where $\rho$ is near-diagonal.}
\end{figure}

While these procedures are theoretically sound, in practice there are StN problems. This can be seen from Eq.~(\ref{eq:expectation}) --- at nonzero $\beta$, $\rho_{ij}$ is generally nonzero between any $\Psi_i$ and $\Psi_j$. Therefore $\langle \delta_{ij}\rangle_{\rho}=\sum_i \rho_{ii}/\sum_{i,j} \rho_{ij}\propto e^{-V}$ where $V$ is the spatial volume. This is itself not a problem: this overall normalization can be disregarded or measured efficiently as discussed in the appendix. However, the physical expectation value $\langle \mathcal O(t)\rangle$ should approach a constant value in the infinite-volume limit, which implies that $\langle \mathcal O(t)\rangle_\rho$ must itself be exponentially small in the volume. As long as typical $O(t)_{ji}$ are not exponentially small in the volume, this will constitute a StN problem.

At short times, this StN problem is due to $\mathcal O(t)$ having few non-vanishing matrix elements --- an overlap problem between $O(t)_{ij}$ and $\rho_{ji}$. For an observable chosen such that $\mathcal O(0)$ is diagonal, there are $\sim e^{V}$ non-vanishing elements, out of $\sim e^{2V}$ total. The nonvanishing $\rho_{ij}$ are thus sampled exponentially infrequently by the OBC Monte Carlo.  At longer times, the generic matrix element is nonzero, and the StN problem arises instead from fine cancellations between positive and negative elements. 

We demonstrate this by computing the $\mathrm{StN}(t)$ of $\mathcal O(t)$, defined as the ratio of the mean to the standard deviation of the random variable $\mathcal O(t)_{ij}$ sampled from $\rho_{ij}/\sum_{k,\ell} \rho_{k\ell}$.   For this example, the one-dimensional ferromagnetic Heisenberg model $H=-J \sum_{i,\alpha} \sigma^\alpha_i\otimes \sigma^\alpha_{i+1}$ with coupling $J=1$ is simulated with $N \in [4,12]$ sites.  The $\mathrm{StN}(t)$ of the spin-spin correlator $\mathcal O(t)=\sigma^z_1(0)\sigma^z_1(t)$ acting on one spin, averaged over early and late times, is in Fig.~\ref{fig:stn}.  

A variety of methods can reduce the StN in practice. In DMQMC, the StN problem requires exponentially large samplings to sufficiently populate $\rho$~\cite{PhysRevB.89.245124}.  To mitigate these large resource requirements, the authors reweighted, reducing the probability of sampling far from the diagonal of $\rho$.  This was sufficient for the time-independent $S=1/2$ antiferromagnetic Heisenberg model. We expect an analogous method is effective in countering the short-time StN problem, where $\mathcal O(t)_{ij}$ vanishes far from the diagonal.

The severity of the StN problem is basis-dependent, which suggests that performing these calculations in a different basis can alleviate the issue as well.  In the eigenbasis of $H$, $\rho$ is diagonal and $\langle \delta_{ij}\rangle_{\rho}=1$.  One illustrative example is that of free fields.  For a free theory, the eigenbasis of the Hamiltonian is the momentum basis, and therefore computations in this basis will have no StN problem. For strongly-coupled theories, finding the eigenbasis is unlikely, but physical intuition may motivate bases with better scaling with reduced noise.

Another path to alleviating the StN problem is reducing the dimension of $\rho$.  To do so without affecting the IR, we must wisely remove states from $\rho$.  For gauge theories, the number of group elements $n$ can be infinite (e.g. $SU(3)$) or merely gigantic (e.g. discrete groups like $\ds$, suggested to approximate $SU(3)$~\cite{Alexandru:2019nsa}).   In this case, a pure gauge lattice has $n^V$ states in the computational basis and the dimension of $\rho$ is $n^{2V}$, with some fraction gauge-invariant.  The StN is naively $\langle \delta_{ij}\rangle_{\rho}\approx \frac{n^V}{n^{2V}}$.  However, this large density of states is dominated by UV lattice artifacts,  particularly in $(3+1)$d, which do not contribute to IR physics. Thus, these states can be elided without altering physical measurements.

\emph{Smearing} methods use projection operators to smooth the field configurations in lattice QCD.  These can be thought of as space or spacetime averaging for gauge theories which removes UV lattice artifacts, reducing the number of states to $n_{eff}<n$. Thus $\rho\rightarrow\rho_{eff}$, the size is reduced to $n_{eff}^{2V}$, and the StN improves to $\langle \delta_{ij}\rangle_{\rho_{eff}}\approx {n_{eff}^{-V}}$.  For gauge fields, smearing methods include \emph{APE}~\cite{Albanese:1987ds}, \emph{HYP}~\cite{Hasenfratz:2001hp}, and \emph{Stout}~\cite{Morningstar:2003gk}.  For fermionic fields, techniques include \emph{Wuppertal}~\cite{Gusken:1989ad} and \emph{Jacobi} smearing~\cite{Allton:1993wc}, which the free fermion kinetic energy to smooth.  Finally, the recent development in momentum smearing for moving hadrons~\cite{Bali:2016lva,Wu:2018tvt} will prove useful for extracting scattering data from a quantum processor.

Low-lying physical states also contribute to the StN problem.  Parisi~\cite{Parisi:1983ae} and Lepage~\cite{Lepage:1989hd} recognized that while $|\psi\rangle$ is the lowest state in $\langle P^\dag(\tau)P(0)\rangle$, it need not be lowest for the variance, estimated by $\langle P^\dag(\tau) P(0)P^\dag(0)P(\tau)\rangle$.  For example, the proton, $p$, is the lowest state overlapping with a 3 quark $P_{3q}$ operator. For the variance, the lowest state is instead 3 pions, $3\pi$. This implies that the StN for proton observables scales $\propto e^{-(E_{p}-\frac{3}{2}E_{\pi}) \tau}$ at long $\tau$.  Together with excited states, this limits the $\tau$ range to a ``golden window'' where properties can be robustly extracted.  Practitioners extend this window by another method, \emph{improved interpolators}, which removes excited states, variance-overlapping states, or both.

The idea of improved interpolators is to build projection operators $\tilde{P}$ from multiple $P$ having the correct quantum numbers and spatial distributions to improve state overlap.  This was done early on variationally~\cite{Michael:1985ne} and recently with more complex methods like distillation~\cite{Peardon:2009gh}.  Distillation has aided the calculation of a wide range of observables: multi-hadron states~\cite{Culver:2019vvu}, excited state spectra~\cite{Liu:2012ze}, coupled-channel resonances~\cite{Woss:2019hse}, and nucleon charges~\cite{Egerer:2018xgu}.  In a few cases, improved interpolators directly addressed StN issues~\cite{Beane:2009gs,Wagman:2016bam,Wagman:2017xfh}; we highlight the intuitive discussion in~\cite{Detmold:2014rfa}.  These interpolators are all usable with our method.  Quantum processors may also prove useful in finding improved interpolators~\cite{Avkhadiev:2019niu}.

Put together, these techniques reduce the StN problem of our method.  This is done by including smearing $\Box$ or improved interpolators $\tilde P$ leading to
\begin{equation}
\left<\psi|\mathcal O(t)|\psi\right> = \frac{\Tr \tilde{P}^\dag\Box^\dag e^{-\beta H_0}\Box\tilde{P} e^{i H_1 t} \mathcal O e^{-i H_1 t}}{\Tr e^{-\beta H_0}}.
\end{equation} 
Since the StN scales like $e^{-V}$, for large $V$, this may prove necessary, although for problems amenable to current resources, its use is limited.  One may also imagine smeared $H_1$ at the expense of circuit depth to reduce errors.

\begin{figure}
\vspace{-0.2cm}
\begin{center}
\begin{picture}(250,130)
\put(0,0){\includegraphics[width=0.78\linewidth]{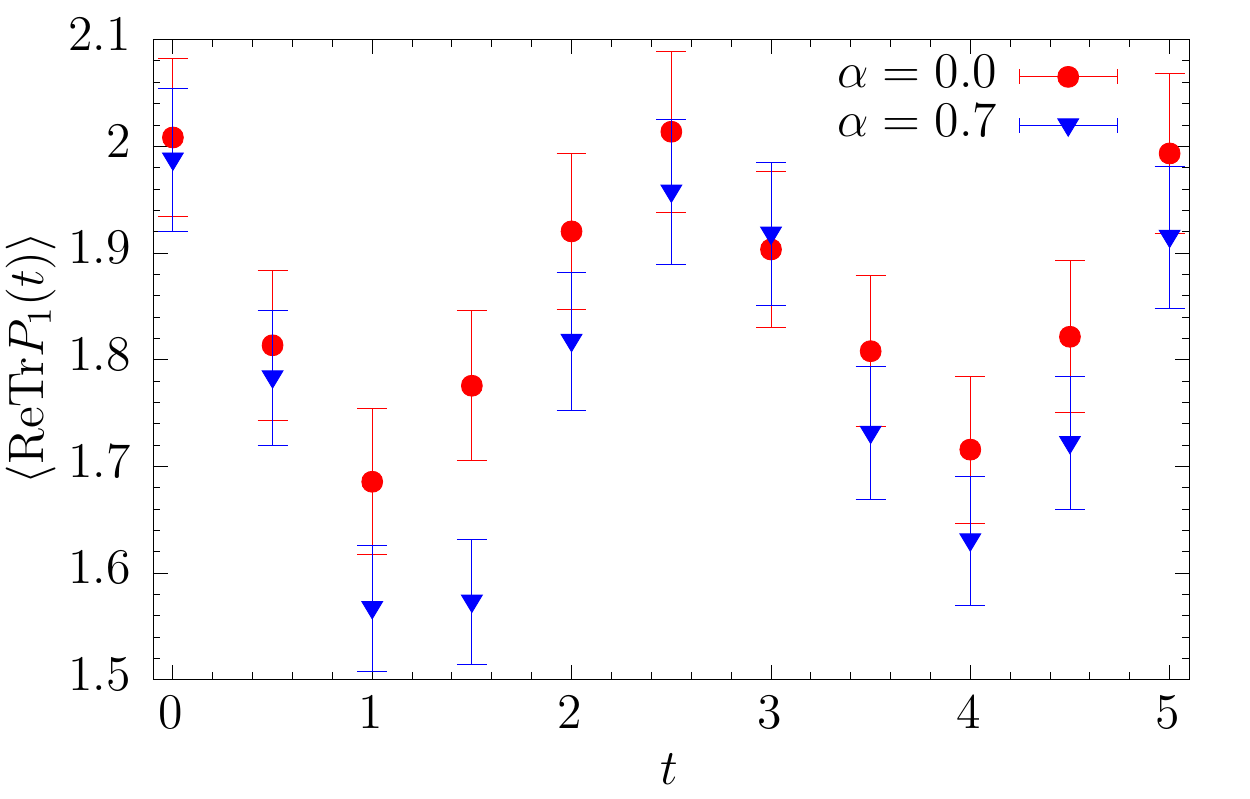}}
\put(190,21){\includegraphics[width=0.2\linewidth]{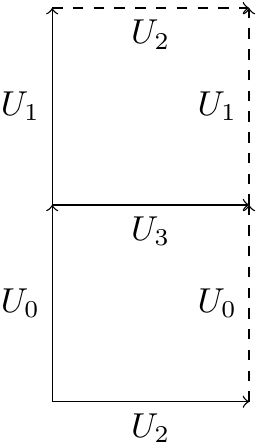}}
\end{picture}
\end{center}
\vspace{-0.5cm}
\caption{\label{fig:d4-results}(left) Expectation value of $\Re\Tr P_1=U_2^\dagger U_0^\dagger U_3 U_0$ vs. $t$ from a quantum simulator. In blue (red) is the result for $\beta_W=0.9$ with(out) smearing $\alpha=0.7$. (right) Spatial lattice where dashed lines indicate repeated links due to PBC.}
\end{figure}

As a demonstration, we simulate the $D_4$ gauge theory with the Wilson action of~\cite{Lamm:2019bik} (right of Fig.~\ref{fig:d4-results}), which needs 12 qubits for the $8^4$ states and 2 ancillary qubits.  The Euclidean calculation had $N_\tau=10$ and Wilson coupling $\beta_W=0.9$, chosen near the $\beta_W\rightarrow\infty$ ground state with the gauge-invariant projection of all $U_i=\id$. We sampled $2 \times 10^4$ configurations separated by 10 steps.  The gauge-invariant states representing the $\Psi_i$ and $\Psi_j$ were passed to a noise-free quantum simulator.  Time evolution is performed with trotterization step $\delta t=0.5$ and $\beta_W=0.8$. The expectation value of one plaquette is measured versus $t$ (Fig.~\ref{fig:d4-results}).  The real-time code is written in \texttt{qiskit}~\cite{santos2017ibm} and simulated for 10 shots per configuration.  In additional to the uncertainty from StN, other statistical uncertainties are the classical Monte Carlo variance and the finite quantum shots.  We estimate one systematic, finite $\delta t$, by the maximum deviation between the exact and trotterize $\beta_W\rightarrow \infty$ results of~\cite{Lamm:2019bik}.  The error budget is in Table~\ref{tab:errors}.

\begin{table}
\caption{\label{tab:errors}Summary of the fractional uncertainties in determination of $\langle \Re\Tr P_1 (t)\rangle$}
\begin{center}
\begin{tabular}
{c c c}
\hline\hline
Source & $\alpha=0.0$ & $\alpha=0.7$\\
\hline
Classical Statistics &2.3\%&2.0\%\\
Quantum Shots &1.3\%&1.0\%\\
$\langle \delta_{ij}\rangle_{\rho}$&1.8\% &1.6\%\\
Trotterization &2.5\%&2.5\%\\
\hline\hline
\end{tabular}
\end{center}
\end{table}

We also perform $APE$ smearing to improve the StN.  Due to the small lattice, we only smear link $U_3$ with $\alpha=0.7$.  Agreement is found with the unsmeared results. The average number of $D_4$ elements that $U_3$ samples goes from $n=3.08(2)$ to $n_{eff}=2.98(2)$.  $\langle \delta_{ij} \rangle_\rho$ increases from 7.8\% to 10.0\%. From this, we naively estimate an increased StN of $(3.08/2.98)^V$ for larger lattices.

The reduction in circuit depth is a crucial advantage of this procedure. Adiabatic state preparation requires repeated time evolution, requiring $\sim 1000$ gates \emph{per step} for the $D_4$ model. By contrast, preparing $\Psi_i$ requires only $\sim 1000$ gates, a substantial savings. For larger lattices, the preparation cost for $\Psi_i$ is expected to be $O(V)$, whereas adiabatic methods are generally $O(V^{4/3})$ for a 3d lattice.

We have proposed a method for computing real-time observables without performing the expensive step of preparing the highly-entangled state on the quantum processor.  On current quantum processors, this is a worthy trade-off. To achieve this, we use standard lattice methods to simulate the system at low temperature, and use the quantum processor as a black box to compute the matrix elements between basis states.  While this method will have issues with signal to noise, a number of techniques for mitigating can be applied, which may prove manageable in the NISQ era. Theories with matter fields integrated out on the Euclidean lattice (i.e. fermions) are the critical next step in this line of work.  Future work could use this method to investigate properties like PDFs~\cite{Lamm:2019uyc} or domain wall dynamics~\cite{Tan:2019kya}.

\begin{acknowledgments}
H.L. is supported by a Department of Energy QuantiSED grant.  S.H., H.L, and S.L., are supported by the U.S. Department of Energy under Contract No.~DE-FG02-93ER-40762. We are grateful to Tom Cohen, Paulo Bedaque, Colin Egerer, Raju Venugopalan, and Michael Wagman for numerous discussions helpful to developing this work. Fermilab is operated by Fermi Research Alliance, LLC under contract number DE-AC02-07CH11359 with the United States Department of Energy. The authors acknowledge the University of Maryland supercomputing resources (http://hpcc.umd.edu) made available for conducting the research reported in this paper. 
\end{acknowledgments}


\bibliographystyle{apsrev4-1}
\bibliography{wise}

\begin{thebibliography}{54}%
\makeatletter
\providecommand \@ifxundefined [1]{%
 \@ifx{#1\undefined}
}%
\providecommand \@ifnum [1]{%
 \ifnum #1\expandafter \@firstoftwo
 \else \expandafter \@secondoftwo
 \fi
}%
\providecommand \@ifx [1]{%
 \ifx #1\expandafter \@firstoftwo
 \else \expandafter \@secondoftwo
 \fi
}%
\providecommand \natexlab [1]{#1}%
\providecommand \enquote  [1]{``#1''}%
\providecommand \bibnamefont  [1]{#1}%
\providecommand \bibfnamefont [1]{#1}%
\providecommand \citenamefont [1]{#1}%
\providecommand \href@noop [0]{\@secondoftwo}%
\providecommand \href [0]{\begingroup \@sanitize@url \@href}%
\providecommand \@href[1]{\@@startlink{#1}\@@href}%
\providecommand \@@href[1]{\endgroup#1\@@endlink}%
\providecommand \@sanitize@url [0]{\catcode `\\12\catcode `\$12\catcode
  `\&12\catcode `\#12\catcode `\^12\catcode `\_12\catcode `\%12\relax}%
\providecommand \@@startlink[1]{}%
\providecommand \@@endlink[0]{}%
\providecommand \url  [0]{\begingroup\@sanitize@url \@url }%
\providecommand \@url [1]{\endgroup\@href {#1}{\urlprefix }}%
\providecommand \urlprefix  [0]{URL }%
\providecommand \Eprint [0]{\href }%
\providecommand \doibase [0]{http://dx.doi.org/}%
\providecommand \selectlanguage [0]{\@gobble}%
\providecommand \bibinfo  [0]{\@secondoftwo}%
\providecommand \bibfield  [0]{\@secondoftwo}%
\providecommand \translation [1]{[#1]}%
\providecommand \BibitemOpen [0]{}%
\providecommand \bibitemStop [0]{}%
\providecommand \bibitemNoStop [0]{.\EOS\space}%
\providecommand \EOS [0]{\spacefactor3000\relax}%
\providecommand \BibitemShut  [1]{\csname bibitem#1\endcsname}%
\let\auto@bib@innerbib\@empty
\bibitem [{\citenamefont {Alexandru}\ \emph {et~al.}(2016)\citenamefont
  {Alexandru}, \citenamefont {Basar}, \citenamefont {Bedaque}, \citenamefont
  {Vartak},\ and\ \citenamefont {Warrington}}]{Alexandru:2016gsd}%
  \BibitemOpen
  \bibfield  {author} {\bibinfo {author} {\bibfnamefont {A.}~\bibnamefont
  {Alexandru}}, \bibinfo {author} {\bibfnamefont {G.}~\bibnamefont {Basar}},
  \bibinfo {author} {\bibfnamefont {P.~F.}\ \bibnamefont {Bedaque}}, \bibinfo
  {author} {\bibfnamefont {S.}~\bibnamefont {Vartak}}, \ and\ \bibinfo {author}
  {\bibfnamefont {N.~C.}\ \bibnamefont {Warrington}},\ }\href {\doibase
  10.1103/PhysRevLett.117.081602} {\bibfield  {journal} {\bibinfo  {journal}
  {Phys. Rev. Lett.}\ }\textbf {\bibinfo {volume} {117}},\ \bibinfo {pages}
  {081602} (\bibinfo {year} {2016})},\ \Eprint
  {http://arxiv.org/abs/1605.08040} {arXiv:1605.08040 [hep-lat]} \BibitemShut
  {NoStop}%
\bibitem [{\citenamefont {Lloyd}(1996)}]{Lloyd1073}%
  \BibitemOpen
  \bibfield  {author} {\bibinfo {author} {\bibfnamefont {S.}~\bibnamefont
  {Lloyd}},\ }\href {\doibase 10.1126/science.273.5278.1073} {\bibfield
  {journal} {\bibinfo  {journal} {Science}\ }\textbf {\bibinfo {volume}
  {273}},\ \bibinfo {pages} {1073} (\bibinfo {year} {1996})}\BibitemShut
  {NoStop}%
\bibitem [{\citenamefont {Ortiz}\ \emph {et~al.}(2001)\citenamefont {Ortiz},
  \citenamefont {Gubernatis}, \citenamefont {Knill},\ and\ \citenamefont
  {Laflamme}}]{Ortiz:2000gc}%
  \BibitemOpen
  \bibfield  {author} {\bibinfo {author} {\bibfnamefont {G.}~\bibnamefont
  {Ortiz}}, \bibinfo {author} {\bibfnamefont {J.~E.}\ \bibnamefont
  {Gubernatis}}, \bibinfo {author} {\bibfnamefont {E.}~\bibnamefont {Knill}}, \
  and\ \bibinfo {author} {\bibfnamefont {R.}~\bibnamefont {Laflamme}},\ }\href
  {\doibase 10.1103/PhysRevA.64.022319} {\bibfield  {journal} {\bibinfo
  {journal} {Phys. Rev.}\ }\textbf {\bibinfo {volume} {A64}},\ \bibinfo {pages}
  {022319} (\bibinfo {year} {2001})},\ \Eprint
  {http://arxiv.org/abs/cond-mat/0012334} {arXiv:cond-mat/0012334 [cond-mat]}
  \BibitemShut {NoStop}%
\bibitem [{\citenamefont {Zohar}\ \emph {et~al.}(2017)\citenamefont {Zohar},
  \citenamefont {Farace}, \citenamefont {Reznik},\ and\ \citenamefont
  {Cirac}}]{Zohar:2016iic}%
  \BibitemOpen
  \bibfield  {author} {\bibinfo {author} {\bibfnamefont {E.}~\bibnamefont
  {Zohar}}, \bibinfo {author} {\bibfnamefont {A.}~\bibnamefont {Farace}},
  \bibinfo {author} {\bibfnamefont {B.}~\bibnamefont {Reznik}}, \ and\ \bibinfo
  {author} {\bibfnamefont {J.~I.}\ \bibnamefont {Cirac}},\ }\href {\doibase
  10.1103/PhysRevA.95.023604} {\bibfield  {journal} {\bibinfo  {journal} {Phys.
  Rev.}\ }\textbf {\bibinfo {volume} {A95}},\ \bibinfo {pages} {023604}
  (\bibinfo {year} {2017})},\ \Eprint {http://arxiv.org/abs/1607.08121}
  {arXiv:1607.08121 [quant-ph]} \BibitemShut {NoStop}%
\bibitem [{\citenamefont {Lamm}\ \emph
  {et~al.}(2019{\natexlab{a}})\citenamefont {Lamm}, \citenamefont {Lawrence},\
  and\ \citenamefont {Yamauchi}}]{Lamm:2019bik}%
  \BibitemOpen
  \bibfield  {author} {\bibinfo {author} {\bibfnamefont {H.}~\bibnamefont
  {Lamm}}, \bibinfo {author} {\bibfnamefont {S.}~\bibnamefont {Lawrence}}, \
  and\ \bibinfo {author} {\bibfnamefont {Y.}~\bibnamefont {Yamauchi}} (\bibinfo
  {collaboration} {NuQS}),\ }\href {\doibase 10.1103/PhysRevD.100.034518}
  {\bibfield  {journal} {\bibinfo  {journal} {Phys. Rev.}\ }\textbf {\bibinfo
  {volume} {D100}},\ \bibinfo {pages} {034518} (\bibinfo {year}
  {2019}{\natexlab{a}})},\ \Eprint {http://arxiv.org/abs/1903.08807}
  {arXiv:1903.08807 [hep-lat]} \BibitemShut {NoStop}%
\bibitem [{\citenamefont {Klco}\ \emph {et~al.}()\citenamefont {Klco},
  \citenamefont {Stryker},\ and\ \citenamefont {Savage}}]{Klco:2019evd}%
  \BibitemOpen
  \bibfield  {author} {\bibinfo {author} {\bibfnamefont {N.}~\bibnamefont
  {Klco}}, \bibinfo {author} {\bibfnamefont {J.~R.}\ \bibnamefont {Stryker}}, \
  and\ \bibinfo {author} {\bibfnamefont {M.~J.}\ \bibnamefont {Savage}},\
  }\href@noop {} {\ }\Eprint {http://arxiv.org/abs/1908.06935}
  {arXiv:1908.06935 [quant-ph]} \BibitemShut {NoStop}%
\bibitem [{\citenamefont {Jordan}\ \emph {et~al.}(2012)\citenamefont {Jordan},
  \citenamefont {Lee},\ and\ \citenamefont {Preskill}}]{Jordan:2011ne}%
  \BibitemOpen
  \bibfield  {author} {\bibinfo {author} {\bibfnamefont {S.~P.}\ \bibnamefont
  {Jordan}}, \bibinfo {author} {\bibfnamefont {K.~S.~M.}\ \bibnamefont {Lee}},
  \ and\ \bibinfo {author} {\bibfnamefont {J.}~\bibnamefont {Preskill}},\
  }\href {\doibase 10.1126/science.1217069} {\bibfield  {journal} {\bibinfo
  {journal} {Science}\ }\textbf {\bibinfo {volume} {336}},\ \bibinfo {pages}
  {1130} (\bibinfo {year} {2012})},\ \Eprint {http://arxiv.org/abs/1111.3633}
  {arXiv:1111.3633 [quant-ph]} \BibitemShut {NoStop}%
\bibitem [{\citenamefont {Jordan}\ \emph {et~al.}(2011)\citenamefont {Jordan},
  \citenamefont {Lee},\ and\ \citenamefont {Preskill}}]{Jordan:2011ci}%
  \BibitemOpen
  \bibfield  {author} {\bibinfo {author} {\bibfnamefont {S.~P.}\ \bibnamefont
  {Jordan}}, \bibinfo {author} {\bibfnamefont {K.~S.~M.}\ \bibnamefont {Lee}},
  \ and\ \bibinfo {author} {\bibfnamefont {J.}~\bibnamefont {Preskill}},\
  }\href@noop {} {\  (\bibinfo {year} {2011})},\ \bibinfo {note} {[Quant. Inf.
  Comput.14,1014(2014)]},\ \Eprint {http://arxiv.org/abs/1112.4833}
  {arXiv:1112.4833 [hep-th]} \BibitemShut {NoStop}%
\bibitem [{\citenamefont {García-Álvarez}\ \emph {et~al.}(2015)\citenamefont
  {García-Álvarez}, \citenamefont {Casanova}, \citenamefont {Mezzacapo},
  \citenamefont {Egusquiza}, \citenamefont {Lamata}, \citenamefont {Romero},\
  and\ \citenamefont {Solano}}]{Garcia-Alvarez:2014uda}%
  \BibitemOpen
  \bibfield  {author} {\bibinfo {author} {\bibfnamefont {L.}~\bibnamefont
  {García-Álvarez}}, \bibinfo {author} {\bibfnamefont {J.}~\bibnamefont
  {Casanova}}, \bibinfo {author} {\bibfnamefont {A.}~\bibnamefont {Mezzacapo}},
  \bibinfo {author} {\bibfnamefont {I.~L.}\ \bibnamefont {Egusquiza}}, \bibinfo
  {author} {\bibfnamefont {L.}~\bibnamefont {Lamata}}, \bibinfo {author}
  {\bibfnamefont {G.}~\bibnamefont {Romero}}, \ and\ \bibinfo {author}
  {\bibfnamefont {E.}~\bibnamefont {Solano}},\ }\href {\doibase
  10.1103/PhysRevLett.114.070502} {\bibfield  {journal} {\bibinfo  {journal}
  {Phys. Rev. Lett.}\ }\textbf {\bibinfo {volume} {114}},\ \bibinfo {pages}
  {070502} (\bibinfo {year} {2015})},\ \Eprint {http://arxiv.org/abs/1404.2868}
  {arXiv:1404.2868 [quant-ph]} \BibitemShut {NoStop}%
\bibitem [{\citenamefont {Jordan}\ \emph {et~al.}(2014)\citenamefont {Jordan},
  \citenamefont {Lee},\ and\ \citenamefont {Preskill}}]{Jordan:2014tma}%
  \BibitemOpen
  \bibfield  {author} {\bibinfo {author} {\bibfnamefont {S.~P.}\ \bibnamefont
  {Jordan}}, \bibinfo {author} {\bibfnamefont {K.~S.~M.}\ \bibnamefont {Lee}},
  \ and\ \bibinfo {author} {\bibfnamefont {J.}~\bibnamefont {Preskill}},\
  }\href@noop {} {\  (\bibinfo {year} {2014})},\ \Eprint
  {http://arxiv.org/abs/1404.7115} {arXiv:1404.7115 [hep-th]} \BibitemShut
  {NoStop}%
\bibitem [{\citenamefont {Jordan}\ \emph {et~al.}(2017)\citenamefont {Jordan},
  \citenamefont {Krovi}, \citenamefont {Lee},\ and\ \citenamefont
  {Preskill}}]{Jordan:2017lea}%
  \BibitemOpen
  \bibfield  {author} {\bibinfo {author} {\bibfnamefont {S.~P.}\ \bibnamefont
  {Jordan}}, \bibinfo {author} {\bibfnamefont {H.}~\bibnamefont {Krovi}},
  \bibinfo {author} {\bibfnamefont {K.~S.~M.}\ \bibnamefont {Lee}}, \ and\
  \bibinfo {author} {\bibfnamefont {J.}~\bibnamefont {Preskill}},\ }\href@noop
  {} {\  (\bibinfo {year} {2017})},\ \Eprint {http://arxiv.org/abs/1703.00454}
  {arXiv:1703.00454 [quant-ph]} \BibitemShut {NoStop}%
\bibitem [{\citenamefont {Hamed~Moosavian}\ and\ \citenamefont
  {Jordan}(2018)}]{Moosavian:2017tkv}%
  \BibitemOpen
  \bibfield  {author} {\bibinfo {author} {\bibfnamefont {A.}~\bibnamefont
  {Hamed~Moosavian}}\ and\ \bibinfo {author} {\bibfnamefont {S.}~\bibnamefont
  {Jordan}},\ }\href {\doibase 10.1103/PhysRevA.98.012332} {\bibfield
  {journal} {\bibinfo  {journal} {Phys. Rev.}\ }\textbf {\bibinfo {volume}
  {A98}},\ \bibinfo {pages} {012332} (\bibinfo {year} {2018})},\ \Eprint
  {http://arxiv.org/abs/1711.04006} {arXiv:1711.04006 [quant-ph]} \BibitemShut
  {NoStop}%
\bibitem [{\citenamefont {Moosavian}\ \emph {et~al.}(2019)\citenamefont
  {Moosavian}, \citenamefont {Garrison},\ and\ \citenamefont
  {Jordan}}]{Moosavian:2019rxg}%
  \BibitemOpen
  \bibfield  {author} {\bibinfo {author} {\bibfnamefont {A.~H.}\ \bibnamefont
  {Moosavian}}, \bibinfo {author} {\bibfnamefont {J.~R.}\ \bibnamefont
  {Garrison}}, \ and\ \bibinfo {author} {\bibfnamefont {S.~P.}\ \bibnamefont
  {Jordan}},\ }\href@noop {} {\  (\bibinfo {year} {2019})},\ \Eprint
  {http://arxiv.org/abs/1911.03505} {arXiv:1911.03505 [quant-ph]} \BibitemShut
  {NoStop}%
\bibitem [{\citenamefont {Gustafson}\ \emph {et~al.}()\citenamefont
  {Gustafson}, \citenamefont {Dreher}, \citenamefont {Hang},\ and\
  \citenamefont {Meurice}}]{Gustafson:2019vsd}%
  \BibitemOpen
  \bibfield  {author} {\bibinfo {author} {\bibfnamefont {E.}~\bibnamefont
  {Gustafson}}, \bibinfo {author} {\bibfnamefont {P.}~\bibnamefont {Dreher}},
  \bibinfo {author} {\bibfnamefont {Z.}~\bibnamefont {Hang}}, \ and\ \bibinfo
  {author} {\bibfnamefont {Y.}~\bibnamefont {Meurice}},\ }\href@noop {} {\
  }\Eprint {http://arxiv.org/abs/1910.09478} {arXiv:1910.09478 [hep-lat]}
  \BibitemShut {NoStop}%
\bibitem [{\citenamefont {Kokail}\ \emph {et~al.}(2018)\citenamefont {Kokail}
  \emph {et~al.}}]{Kokail:2018eiw}%
  \BibitemOpen
  \bibfield  {author} {\bibinfo {author} {\bibfnamefont {C.}~\bibnamefont
  {Kokail}} \emph {et~al.},\ }\href@noop {} {\  (\bibinfo {year} {2018})},\
  \Eprint {http://arxiv.org/abs/1810.03421} {arXiv:1810.03421 [quant-ph]}
  \BibitemShut {NoStop}%
\bibitem [{\citenamefont {Lamm}\ and\ \citenamefont
  {Lawrence}(2018)}]{Lamm:2018siq}%
  \BibitemOpen
  \bibfield  {author} {\bibinfo {author} {\bibfnamefont {H.}~\bibnamefont
  {Lamm}}\ and\ \bibinfo {author} {\bibfnamefont {S.}~\bibnamefont
  {Lawrence}},\ }\href {\doibase 10.1103/PhysRevLett.121.170501} {\bibfield
  {journal} {\bibinfo  {journal} {Phys. Rev. Lett.}\ }\textbf {\bibinfo
  {volume} {121}},\ \bibinfo {pages} {170501} (\bibinfo {year} {2018})},\
  \Eprint {http://arxiv.org/abs/1806.06649} {arXiv:1806.06649 [quant-ph]}
  \BibitemShut {NoStop}%
\bibitem [{\citenamefont {Klco}\ and\ \citenamefont
  {Savage}(2019{\natexlab{a}})}]{Klco:2019xro}%
  \BibitemOpen
  \bibfield  {author} {\bibinfo {author} {\bibfnamefont {N.}~\bibnamefont
  {Klco}}\ and\ \bibinfo {author} {\bibfnamefont {M.~J.}\ \bibnamefont
  {Savage}},\ }\href@noop {} {\  (\bibinfo {year} {2019}{\natexlab{a}})},\
  \Eprint {http://arxiv.org/abs/1904.10440} {arXiv:1904.10440 [quant-ph]}
  \BibitemShut {NoStop}%
\bibitem [{\citenamefont {Klco}\ and\ \citenamefont
  {Savage}(2019{\natexlab{b}})}]{Klco:2019yrb}%
  \BibitemOpen
  \bibfield  {author} {\bibinfo {author} {\bibfnamefont {N.}~\bibnamefont
  {Klco}}\ and\ \bibinfo {author} {\bibfnamefont {M.~J.}\ \bibnamefont
  {Savage}},\ }\href@noop {} {\  (\bibinfo {year} {2019}{\natexlab{b}})},\
  \Eprint {http://arxiv.org/abs/1912.03577} {arXiv:1912.03577 [quant-ph]}
  \BibitemShut {NoStop}%
\bibitem [{\citenamefont {Blunt}\ \emph {et~al.}(2014)\citenamefont {Blunt},
  \citenamefont {Rogers}, \citenamefont {Spencer},\ and\ \citenamefont
  {Foulkes}}]{PhysRevB.89.245124}%
  \BibitemOpen
  \bibfield  {author} {\bibinfo {author} {\bibfnamefont {N.~S.}\ \bibnamefont
  {Blunt}}, \bibinfo {author} {\bibfnamefont {T.~W.}\ \bibnamefont {Rogers}},
  \bibinfo {author} {\bibfnamefont {J.~S.}\ \bibnamefont {Spencer}}, \ and\
  \bibinfo {author} {\bibfnamefont {W.~M.~C.}\ \bibnamefont {Foulkes}},\ }\href
  {\doibase 10.1103/PhysRevB.89.245124} {\bibfield  {journal} {\bibinfo
  {journal} {Phys. Rev. B}\ }\textbf {\bibinfo {volume} {89}},\ \bibinfo
  {pages} {245124} (\bibinfo {year} {2014})}\BibitemShut {NoStop}%
\bibitem [{\citenamefont {Parisi}(1984)}]{Parisi:1983ae}%
  \BibitemOpen
  \bibfield  {author} {\bibinfo {author} {\bibfnamefont {G.}~\bibnamefont
  {Parisi}},\ }\href {\doibase 10.1016/0370-1573(84)90081-4} {\bibfield
  {journal} {\bibinfo  {journal} {Phys. Rept.}\ }\textbf {\bibinfo {volume}
  {103}},\ \bibinfo {pages} {203} (\bibinfo {year} {1984})}\BibitemShut
  {NoStop}%
\bibitem [{\citenamefont {Lepage}(1989)}]{Lepage:1989hd}%
  \BibitemOpen
  \bibfield  {author} {\bibinfo {author} {\bibfnamefont {G.~P.}\ \bibnamefont
  {Lepage}},\ }in\ \href@noop {} {\emph {\bibinfo {booktitle} {{Boulder ASI
  1989:97-120}}}}\ (\bibinfo {year} {1989})\ pp.\ \bibinfo {pages}
  {97--120}\BibitemShut {NoStop}%
\bibitem [{\citenamefont {Schwinger}(1961)}]{Schwinger:1960qe}%
  \BibitemOpen
  \bibfield  {author} {\bibinfo {author} {\bibfnamefont {J.~S.}\ \bibnamefont
  {Schwinger}},\ }\href {\doibase 10.1063/1.1703727} {\bibfield  {journal}
  {\bibinfo  {journal} {J. Math. Phys.}\ }\textbf {\bibinfo {volume} {2}},\
  \bibinfo {pages} {407} (\bibinfo {year} {1961})}\BibitemShut {NoStop}%
\bibitem [{\citenamefont {Keldysh}(1964)}]{Keldysh:1964ud}%
  \BibitemOpen
  \bibfield  {author} {\bibinfo {author} {\bibfnamefont {L.~V.}\ \bibnamefont
  {Keldysh}},\ }\href@noop {} {\bibfield  {journal} {\bibinfo  {journal} {Zh.
  Eksp. Teor. Fiz.}\ }\textbf {\bibinfo {volume} {47}},\ \bibinfo {pages}
  {1515} (\bibinfo {year} {1964})},\ \bibinfo {note} {[Sov. Phys.
  JETP20,1018(1965)]}\BibitemShut {NoStop}%
\bibitem [{\citenamefont {Low}\ and\ \citenamefont
  {Chuang}(2017)}]{PhysRevLett.118.010501}%
  \BibitemOpen
  \bibfield  {author} {\bibinfo {author} {\bibfnamefont {G.~H.}\ \bibnamefont
  {Low}}\ and\ \bibinfo {author} {\bibfnamefont {I.~L.}\ \bibnamefont
  {Chuang}},\ }\href {\doibase 10.1103/PhysRevLett.118.010501} {\bibfield
  {journal} {\bibinfo  {journal} {Phys. Rev. Lett.}\ }\textbf {\bibinfo
  {volume} {118}},\ \bibinfo {pages} {010501} (\bibinfo {year}
  {2017})}\BibitemShut {NoStop}%
\bibitem [{\citenamefont {Campbell}(2019)}]{PhysRevLett.123.070503}%
  \BibitemOpen
  \bibfield  {author} {\bibinfo {author} {\bibfnamefont {E.}~\bibnamefont
  {Campbell}},\ }\href {\doibase 10.1103/PhysRevLett.123.070503} {\bibfield
  {journal} {\bibinfo  {journal} {Phys. Rev. Lett.}\ }\textbf {\bibinfo
  {volume} {123}},\ \bibinfo {pages} {070503} (\bibinfo {year}
  {2019})}\BibitemShut {NoStop}%
\bibitem [{\citenamefont {Roggero}\ and\ \citenamefont
  {Carlson}(2019)}]{Roggero:2018hrn}%
  \BibitemOpen
  \bibfield  {author} {\bibinfo {author} {\bibfnamefont {A.}~\bibnamefont
  {Roggero}}\ and\ \bibinfo {author} {\bibfnamefont {J.}~\bibnamefont
  {Carlson}},\ }\href {\doibase 10.1103/PhysRevC.100.034610} {\bibfield
  {journal} {\bibinfo  {journal} {Phys. Rev.}\ }\textbf {\bibinfo {volume}
  {C100}},\ \bibinfo {pages} {034610} (\bibinfo {year} {2019})},\ \Eprint
  {http://arxiv.org/abs/1804.01505} {arXiv:1804.01505 [quant-ph]} \BibitemShut
  {NoStop}%
\bibitem [{\citenamefont {Zohar}\ and\ \citenamefont
  {Cirac}(2018)}]{Zohar:2018cwb}%
  \BibitemOpen
  \bibfield  {author} {\bibinfo {author} {\bibfnamefont {E.}~\bibnamefont
  {Zohar}}\ and\ \bibinfo {author} {\bibfnamefont {J.~I.}\ \bibnamefont
  {Cirac}},\ }\href {\doibase 10.1103/PhysRevB.98.075119} {\bibfield  {journal}
  {\bibinfo  {journal} {Phys. Rev.}\ }\textbf {\bibinfo {volume} {B98}},\
  \bibinfo {pages} {075119} (\bibinfo {year} {2018})},\ \Eprint
  {http://arxiv.org/abs/1805.05347} {arXiv:1805.05347 [quant-ph]} \BibitemShut
  {NoStop}%
\bibitem [{\citenamefont {Clemente}\ \emph {et~al.}()\citenamefont {Clemente}
  \emph {et~al.}}]{Clemente:2020lpr}%
  \BibitemOpen
  \bibfield  {author} {\bibinfo {author} {\bibfnamefont {G.}~\bibnamefont
  {Clemente}} \emph {et~al.},\ }\href@noop {} {\ }\Eprint
  {http://arxiv.org/abs/2001.05328} {arXiv:2001.05328 [hep-lat]} \BibitemShut
  {NoStop}%
\bibitem [{\citenamefont {Luscher}\ and\ \citenamefont
  {Schaefer}(2011)}]{Luscher:2011kk}%
  \BibitemOpen
  \bibfield  {author} {\bibinfo {author} {\bibfnamefont {M.}~\bibnamefont
  {Luscher}}\ and\ \bibinfo {author} {\bibfnamefont {S.}~\bibnamefont
  {Schaefer}},\ }\href {\doibase 10.1007/JHEP07(2011)036} {\bibfield  {journal}
  {\bibinfo  {journal} {JHEP}\ }\textbf {\bibinfo {volume} {07}},\ \bibinfo
  {pages} {036} (\bibinfo {year} {2011})},\ \Eprint
  {http://arxiv.org/abs/1105.4749} {arXiv:1105.4749 [hep-lat]} \BibitemShut
  {NoStop}%
\bibitem [{\citenamefont {Luscher}\ and\ \citenamefont
  {Schaefer}(2013)}]{Luscher:2012av}%
  \BibitemOpen
  \bibfield  {author} {\bibinfo {author} {\bibfnamefont {M.}~\bibnamefont
  {Luscher}}\ and\ \bibinfo {author} {\bibfnamefont {S.}~\bibnamefont
  {Schaefer}},\ }\href {\doibase 10.1016/j.cpc.2012.10.003} {\bibfield
  {journal} {\bibinfo  {journal} {Comput.Phys.Commun.}\ }\textbf {\bibinfo
  {volume} {184}},\ \bibinfo {pages} {519} (\bibinfo {year} {2013})},\ \Eprint
  {http://arxiv.org/abs/1206.2809} {arXiv:1206.2809 [hep-lat]} \BibitemShut
  {NoStop}%
\bibitem [{\citenamefont {H{\"o}llwieser}\ \emph {et~al.}(2019)\citenamefont
  {H{\"o}llwieser}, \citenamefont {Knechtli},\ and\ \citenamefont
  {Korzec}}]{Hollwieser:2019kuc}%
  \BibitemOpen
  \bibfield  {author} {\bibinfo {author} {\bibfnamefont {R.}~\bibnamefont
  {H{\"o}llwieser}}, \bibinfo {author} {\bibfnamefont {F.}~\bibnamefont
  {Knechtli}}, \ and\ \bibinfo {author} {\bibfnamefont {T.}~\bibnamefont
  {Korzec}},\ }\href@noop {} {\  (\bibinfo {year} {2019})},\ \Eprint
  {http://arxiv.org/abs/1907.04309} {arXiv:1907.04309 [hep-lat]} \BibitemShut
  {NoStop}%
\bibitem [{\citenamefont {Florio}\ \emph {et~al.}()\citenamefont {Florio},
  \citenamefont {Kaczmarek},\ and\ \citenamefont {Mazur}}]{Florio:2019nte}%
  \BibitemOpen
  \bibfield  {author} {\bibinfo {author} {\bibfnamefont {A.}~\bibnamefont
  {Florio}}, \bibinfo {author} {\bibfnamefont {O.}~\bibnamefont {Kaczmarek}}, \
  and\ \bibinfo {author} {\bibfnamefont {L.}~\bibnamefont {Mazur}},\
  }\href@noop {} {\ }\Eprint {http://arxiv.org/abs/1903.02894}
  {arXiv:1903.02894 [hep-lat]} \BibitemShut {NoStop}%
\bibitem [{\citenamefont {Alexandru}\ \emph {et~al.}(2019)\citenamefont
  {Alexandru}, \citenamefont {Bedaque}, \citenamefont {Harmalkar},
  \citenamefont {Lamm}, \citenamefont {Lawrence},\ and\ \citenamefont
  {Warrington}}]{Alexandru:2019nsa}%
  \BibitemOpen
  \bibfield  {author} {\bibinfo {author} {\bibfnamefont {A.}~\bibnamefont
  {Alexandru}}, \bibinfo {author} {\bibfnamefont {P.~F.}\ \bibnamefont
  {Bedaque}}, \bibinfo {author} {\bibfnamefont {S.}~\bibnamefont {Harmalkar}},
  \bibinfo {author} {\bibfnamefont {H.}~\bibnamefont {Lamm}}, \bibinfo {author}
  {\bibfnamefont {S.}~\bibnamefont {Lawrence}}, \ and\ \bibinfo {author}
  {\bibfnamefont {N.~C.}\ \bibnamefont {Warrington}} (\bibinfo {collaboration}
  {NuQS}),\ }\href {\doibase 10.1103/PhysRevD.100.114501} {\bibfield  {journal}
  {\bibinfo  {journal} {Phys.Rev.D}\ }\textbf {\bibinfo {volume} {100}},\
  \bibinfo {pages} {114501} (\bibinfo {year} {2019})},\ \Eprint
  {http://arxiv.org/abs/1906.11213} {arXiv:1906.11213 [hep-lat]} \BibitemShut
  {NoStop}%
\bibitem [{\citenamefont {Albanese}\ \emph {et~al.}(1987)\citenamefont
  {Albanese} \emph {et~al.}}]{Albanese:1987ds}%
  \BibitemOpen
  \bibfield  {author} {\bibinfo {author} {\bibfnamefont {M.}~\bibnamefont
  {Albanese}} \emph {et~al.} (\bibinfo {collaboration} {APE}),\ }\href
  {\doibase 10.1016/0370-2693(87)91160-9} {\bibfield  {journal} {\bibinfo
  {journal} {Phys. Lett.}\ }\textbf {\bibinfo {volume} {B192}},\ \bibinfo
  {pages} {163} (\bibinfo {year} {1987})}\BibitemShut {NoStop}%
\bibitem [{\citenamefont {Hasenfratz}\ and\ \citenamefont
  {Knechtli}(2001)}]{Hasenfratz:2001hp}%
  \BibitemOpen
  \bibfield  {author} {\bibinfo {author} {\bibfnamefont {A.}~\bibnamefont
  {Hasenfratz}}\ and\ \bibinfo {author} {\bibfnamefont {F.}~\bibnamefont
  {Knechtli}},\ }\href {\doibase 10.1103/PhysRevD.64.034504} {\bibfield
  {journal} {\bibinfo  {journal} {Phys. Rev.}\ }\textbf {\bibinfo {volume}
  {D64}},\ \bibinfo {pages} {034504} (\bibinfo {year} {2001})},\ \Eprint
  {http://arxiv.org/abs/hep-lat/0103029} {arXiv:hep-lat/0103029 [hep-lat]}
  \BibitemShut {NoStop}%
\bibitem [{\citenamefont {Morningstar}\ and\ \citenamefont
  {Peardon}(2004)}]{Morningstar:2003gk}%
  \BibitemOpen
  \bibfield  {author} {\bibinfo {author} {\bibfnamefont {C.}~\bibnamefont
  {Morningstar}}\ and\ \bibinfo {author} {\bibfnamefont {M.~J.}\ \bibnamefont
  {Peardon}},\ }\href {\doibase 10.1103/PhysRevD.69.054501} {\bibfield
  {journal} {\bibinfo  {journal} {Phys. Rev.}\ }\textbf {\bibinfo {volume}
  {D69}},\ \bibinfo {pages} {054501} (\bibinfo {year} {2004})},\ \Eprint
  {http://arxiv.org/abs/hep-lat/0311018} {arXiv:hep-lat/0311018 [hep-lat]}
  \BibitemShut {NoStop}%
\bibitem [{\citenamefont {Gusken}\ \emph {et~al.}(1989)\citenamefont {Gusken},
  \citenamefont {Low}, \citenamefont {Mutter}, \citenamefont {Sommer},
  \citenamefont {Patel},\ and\ \citenamefont {Schilling}}]{Gusken:1989ad}%
  \BibitemOpen
  \bibfield  {author} {\bibinfo {author} {\bibfnamefont {S.}~\bibnamefont
  {Gusken}}, \bibinfo {author} {\bibfnamefont {U.}~\bibnamefont {Low}},
  \bibinfo {author} {\bibfnamefont {K.~H.}\ \bibnamefont {Mutter}}, \bibinfo
  {author} {\bibfnamefont {R.}~\bibnamefont {Sommer}}, \bibinfo {author}
  {\bibfnamefont {A.}~\bibnamefont {Patel}}, \ and\ \bibinfo {author}
  {\bibfnamefont {K.}~\bibnamefont {Schilling}},\ }\href {\doibase
  10.1016/S0370-2693(89)80034-6} {\bibfield  {journal} {\bibinfo  {journal}
  {Phys. Lett.}\ }\textbf {\bibinfo {volume} {B227}},\ \bibinfo {pages} {266}
  (\bibinfo {year} {1989})}\BibitemShut {NoStop}%
\bibitem [{\citenamefont {Allton}\ \emph {et~al.}(1993)\citenamefont {Allton}
  \emph {et~al.}}]{Allton:1993wc}%
  \BibitemOpen
  \bibfield  {author} {\bibinfo {author} {\bibfnamefont {C.~R.}\ \bibnamefont
  {Allton}} \emph {et~al.} (\bibinfo {collaboration} {UKQCD}),\ }\href
  {\doibase 10.1103/PhysRevD.47.5128} {\bibfield  {journal} {\bibinfo
  {journal} {Phys. Rev.}\ }\textbf {\bibinfo {volume} {D47}},\ \bibinfo {pages}
  {5128} (\bibinfo {year} {1993})},\ \Eprint
  {http://arxiv.org/abs/hep-lat/9303009} {arXiv:hep-lat/9303009 [hep-lat]}
  \BibitemShut {NoStop}%
\bibitem [{\citenamefont {Bali}\ \emph {et~al.}(2016)\citenamefont {Bali},
  \citenamefont {Lang}, \citenamefont {Musch},\ and\ \citenamefont
  {Sch{\"a}fer}}]{Bali:2016lva}%
  \BibitemOpen
  \bibfield  {author} {\bibinfo {author} {\bibfnamefont {G.~S.}\ \bibnamefont
  {Bali}}, \bibinfo {author} {\bibfnamefont {B.}~\bibnamefont {Lang}}, \bibinfo
  {author} {\bibfnamefont {B.~U.}\ \bibnamefont {Musch}}, \ and\ \bibinfo
  {author} {\bibfnamefont {A.}~\bibnamefont {Sch{\"a}fer}},\ }\href {\doibase
  10.1103/PhysRevD.93.094515} {\bibfield  {journal} {\bibinfo  {journal} {Phys.
  Rev.}\ }\textbf {\bibinfo {volume} {D93}},\ \bibinfo {pages} {094515}
  (\bibinfo {year} {2016})},\ \Eprint {http://arxiv.org/abs/1602.05525}
  {arXiv:1602.05525 [hep-lat]} \BibitemShut {NoStop}%
\bibitem [{\citenamefont {Wu}\ \emph {et~al.}(2018)\citenamefont {Wu},
  \citenamefont {Kamleh}, \citenamefont {Leinweber}, \citenamefont {Young},\
  and\ \citenamefont {Zanotti}}]{Wu:2018tvt}%
  \BibitemOpen
  \bibfield  {author} {\bibinfo {author} {\bibfnamefont {J.~J.}\ \bibnamefont
  {Wu}}, \bibinfo {author} {\bibfnamefont {W.}~\bibnamefont {Kamleh}}, \bibinfo
  {author} {\bibfnamefont {D.~B.}\ \bibnamefont {Leinweber}}, \bibinfo {author}
  {\bibfnamefont {R.~D.}\ \bibnamefont {Young}}, \ and\ \bibinfo {author}
  {\bibfnamefont {J.~M.}\ \bibnamefont {Zanotti}},\ }\href {\doibase
  10.1088/1361-6471/aaeb9e} {\bibfield  {journal} {\bibinfo  {journal} {J.
  Phys.}\ }\textbf {\bibinfo {volume} {G45}},\ \bibinfo {pages} {125102}
  (\bibinfo {year} {2018})},\ \Eprint {http://arxiv.org/abs/1807.09429}
  {arXiv:1807.09429 [hep-lat]} \BibitemShut {NoStop}%
\bibitem [{\citenamefont {Michael}(1985)}]{Michael:1985ne}%
  \BibitemOpen
  \bibfield  {author} {\bibinfo {author} {\bibfnamefont {C.}~\bibnamefont
  {Michael}},\ }\href {\doibase 10.1016/0550-3213(85)90297-4} {\bibfield
  {journal} {\bibinfo  {journal} {Nucl. Phys.}\ }\textbf {\bibinfo {volume}
  {B259}},\ \bibinfo {pages} {58} (\bibinfo {year} {1985})}\BibitemShut
  {NoStop}%
\bibitem [{\citenamefont {Peardon}\ \emph {et~al.}(2009)\citenamefont
  {Peardon}, \citenamefont {Bulava}, \citenamefont {Foley}, \citenamefont
  {Morningstar}, \citenamefont {Dudek}, \citenamefont {Edwards}, \citenamefont
  {Joo}, \citenamefont {Lin}, \citenamefont {Richards},\ and\ \citenamefont
  {Juge}}]{Peardon:2009gh}%
  \BibitemOpen
  \bibfield  {author} {\bibinfo {author} {\bibfnamefont {M.}~\bibnamefont
  {Peardon}}, \bibinfo {author} {\bibfnamefont {J.}~\bibnamefont {Bulava}},
  \bibinfo {author} {\bibfnamefont {J.}~\bibnamefont {Foley}}, \bibinfo
  {author} {\bibfnamefont {C.}~\bibnamefont {Morningstar}}, \bibinfo {author}
  {\bibfnamefont {J.}~\bibnamefont {Dudek}}, \bibinfo {author} {\bibfnamefont
  {R.~G.}\ \bibnamefont {Edwards}}, \bibinfo {author} {\bibfnamefont
  {B.}~\bibnamefont {Joo}}, \bibinfo {author} {\bibfnamefont {H.-W.}\
  \bibnamefont {Lin}}, \bibinfo {author} {\bibfnamefont {D.~G.}\ \bibnamefont
  {Richards}}, \ and\ \bibinfo {author} {\bibfnamefont {K.~J.}\ \bibnamefont
  {Juge}} (\bibinfo {collaboration} {Hadron Spectrum}),\ }\href {\doibase
  10.1103/PhysRevD.80.054506} {\bibfield  {journal} {\bibinfo  {journal} {Phys.
  Rev.}\ }\textbf {\bibinfo {volume} {D80}},\ \bibinfo {pages} {054506}
  (\bibinfo {year} {2009})},\ \Eprint {http://arxiv.org/abs/0905.2160}
  {arXiv:0905.2160 [hep-lat]} \BibitemShut {NoStop}%
\bibitem [{\citenamefont {Culver}\ \emph {et~al.}(2019)\citenamefont {Culver},
  \citenamefont {Mai}, \citenamefont {Brett}, \citenamefont {Alexandru},\ and\
  \citenamefont {D{\"o}ring}}]{Culver:2019vvu}%
  \BibitemOpen
  \bibfield  {author} {\bibinfo {author} {\bibfnamefont {C.}~\bibnamefont
  {Culver}}, \bibinfo {author} {\bibfnamefont {M.}~\bibnamefont {Mai}},
  \bibinfo {author} {\bibfnamefont {R.}~\bibnamefont {Brett}}, \bibinfo
  {author} {\bibfnamefont {A.}~\bibnamefont {Alexandru}}, \ and\ \bibinfo
  {author} {\bibfnamefont {M.}~\bibnamefont {D{\"o}ring}},\ }\href@noop {} {\
  (\bibinfo {year} {2019})},\ \Eprint {http://arxiv.org/abs/1911.09047}
  {arXiv:1911.09047 [hep-lat]} \BibitemShut {NoStop}%
\bibitem [{\citenamefont {Liu}\ \emph {et~al.}(2012)\citenamefont {Liu},
  \citenamefont {Moir}, \citenamefont {Peardon}, \citenamefont {Ryan},
  \citenamefont {Thomas}, \citenamefont {Vilaseca}, \citenamefont {Dudek},
  \citenamefont {Edwards}, \citenamefont {Joo},\ and\ \citenamefont
  {Richards}}]{Liu:2012ze}%
  \BibitemOpen
  \bibfield  {author} {\bibinfo {author} {\bibfnamefont {L.}~\bibnamefont
  {Liu}}, \bibinfo {author} {\bibfnamefont {G.}~\bibnamefont {Moir}}, \bibinfo
  {author} {\bibfnamefont {M.}~\bibnamefont {Peardon}}, \bibinfo {author}
  {\bibfnamefont {S.~M.}\ \bibnamefont {Ryan}}, \bibinfo {author}
  {\bibfnamefont {C.~E.}\ \bibnamefont {Thomas}}, \bibinfo {author}
  {\bibfnamefont {P.}~\bibnamefont {Vilaseca}}, \bibinfo {author}
  {\bibfnamefont {J.~J.}\ \bibnamefont {Dudek}}, \bibinfo {author}
  {\bibfnamefont {R.~G.}\ \bibnamefont {Edwards}}, \bibinfo {author}
  {\bibfnamefont {B.}~\bibnamefont {Joo}}, \ and\ \bibinfo {author}
  {\bibfnamefont {D.~G.}\ \bibnamefont {Richards}} (\bibinfo {collaboration}
  {Hadron Spectrum}),\ }\href {\doibase 10.1007/JHEP07(2012)126} {\bibfield
  {journal} {\bibinfo  {journal} {JHEP}\ }\textbf {\bibinfo {volume} {07}},\
  \bibinfo {pages} {126} (\bibinfo {year} {2012})},\ \Eprint
  {http://arxiv.org/abs/1204.5425} {arXiv:1204.5425 [hep-ph]} \BibitemShut
  {NoStop}%
\bibitem [{\citenamefont {Woss}\ \emph {et~al.}(2019)\citenamefont {Woss},
  \citenamefont {Thomas}, \citenamefont {Dudek}, \citenamefont {Edwards},\ and\
  \citenamefont {Wilson}}]{Woss:2019hse}%
  \BibitemOpen
  \bibfield  {author} {\bibinfo {author} {\bibfnamefont {A.~J.}\ \bibnamefont
  {Woss}}, \bibinfo {author} {\bibfnamefont {C.~E.}\ \bibnamefont {Thomas}},
  \bibinfo {author} {\bibfnamefont {J.~J.}\ \bibnamefont {Dudek}}, \bibinfo
  {author} {\bibfnamefont {R.~G.}\ \bibnamefont {Edwards}}, \ and\ \bibinfo
  {author} {\bibfnamefont {D.~J.}\ \bibnamefont {Wilson}},\ }\href {\doibase
  10.1103/PhysRevD.100.054506} {\bibfield  {journal} {\bibinfo  {journal}
  {Phys. Rev.}\ }\textbf {\bibinfo {volume} {D100}},\ \bibinfo {pages} {054506}
  (\bibinfo {year} {2019})},\ \Eprint {http://arxiv.org/abs/1904.04136}
  {arXiv:1904.04136 [hep-lat]} \BibitemShut {NoStop}%
\bibitem [{\citenamefont {Egerer}\ \emph {et~al.}(2019)\citenamefont {Egerer},
  \citenamefont {Richards},\ and\ \citenamefont {Winter}}]{Egerer:2018xgu}%
  \BibitemOpen
  \bibfield  {author} {\bibinfo {author} {\bibfnamefont {C.}~\bibnamefont
  {Egerer}}, \bibinfo {author} {\bibfnamefont {D.}~\bibnamefont {Richards}}, \
  and\ \bibinfo {author} {\bibfnamefont {F.}~\bibnamefont {Winter}},\ }\href
  {\doibase 10.1103/PhysRevD.99.034506} {\bibfield  {journal} {\bibinfo
  {journal} {Phys. Rev.}\ }\textbf {\bibinfo {volume} {D99}},\ \bibinfo {pages}
  {034506} (\bibinfo {year} {2019})},\ \Eprint
  {http://arxiv.org/abs/1810.09991} {arXiv:1810.09991 [hep-lat]} \BibitemShut
  {NoStop}%
\bibitem [{\citenamefont {Beane}\ \emph {et~al.}(2009)\citenamefont {Beane},
  \citenamefont {Detmold}, \citenamefont {Luu}, \citenamefont {Orginos},
  \citenamefont {Parreno}, \citenamefont {Savage}, \citenamefont {Torok},\ and\
  \citenamefont {Walker-Loud}}]{Beane:2009gs}%
  \BibitemOpen
  \bibfield  {author} {\bibinfo {author} {\bibfnamefont {S.~R.}\ \bibnamefont
  {Beane}}, \bibinfo {author} {\bibfnamefont {W.}~\bibnamefont {Detmold}},
  \bibinfo {author} {\bibfnamefont {T.~C.}\ \bibnamefont {Luu}}, \bibinfo
  {author} {\bibfnamefont {K.}~\bibnamefont {Orginos}}, \bibinfo {author}
  {\bibfnamefont {A.}~\bibnamefont {Parreno}}, \bibinfo {author} {\bibfnamefont
  {M.~J.}\ \bibnamefont {Savage}}, \bibinfo {author} {\bibfnamefont
  {A.}~\bibnamefont {Torok}}, \ and\ \bibinfo {author} {\bibfnamefont
  {A.}~\bibnamefont {Walker-Loud}},\ }\href {\doibase
  10.1103/PhysRevD.80.074501} {\bibfield  {journal} {\bibinfo  {journal} {Phys.
  Rev.}\ }\textbf {\bibinfo {volume} {D80}},\ \bibinfo {pages} {074501}
  (\bibinfo {year} {2009})},\ \Eprint {http://arxiv.org/abs/0905.0466}
  {arXiv:0905.0466 [hep-lat]} \BibitemShut {NoStop}%
\bibitem [{\citenamefont {Wagman}\ and\ \citenamefont
  {Savage}(2017{\natexlab{a}})}]{Wagman:2016bam}%
  \BibitemOpen
  \bibfield  {author} {\bibinfo {author} {\bibfnamefont {M.~L.}\ \bibnamefont
  {Wagman}}\ and\ \bibinfo {author} {\bibfnamefont {M.~J.}\ \bibnamefont
  {Savage}},\ }\href {\doibase 10.1103/PhysRevD.96.114508} {\bibfield
  {journal} {\bibinfo  {journal} {Phys. Rev.}\ }\textbf {\bibinfo {volume}
  {D96}},\ \bibinfo {pages} {114508} (\bibinfo {year} {2017}{\natexlab{a}})},\
  \Eprint {http://arxiv.org/abs/1611.07643} {arXiv:1611.07643 [hep-lat]}
  \BibitemShut {NoStop}%
\bibitem [{\citenamefont {Wagman}\ and\ \citenamefont
  {Savage}(2017{\natexlab{b}})}]{Wagman:2017xfh}%
  \BibitemOpen
  \bibfield  {author} {\bibinfo {author} {\bibfnamefont {M.~L.}\ \bibnamefont
  {Wagman}}\ and\ \bibinfo {author} {\bibfnamefont {M.~J.}\ \bibnamefont
  {Savage}},\ }\href@noop {} {\  (\bibinfo {year} {2017}{\natexlab{b}})},\
  \Eprint {http://arxiv.org/abs/1704.07356} {arXiv:1704.07356 [hep-lat]}
  \BibitemShut {NoStop}%
\bibitem [{\citenamefont {Detmold}\ and\ \citenamefont
  {Endres}(2015)}]{Detmold:2014rfa}%
  \BibitemOpen
  \bibfield  {author} {\bibinfo {author} {\bibfnamefont {W.}~\bibnamefont
  {Detmold}}\ and\ \bibinfo {author} {\bibfnamefont {M.~G.}\ \bibnamefont
  {Endres}},\ }\bibfield  {booktitle} {\emph {\bibinfo {booktitle}
  {{Proceedings, 32nd International Symposium on Lattice Field Theory (Lattice
  2014): Brookhaven, NY, USA, June 23-28, 2014}}},\ }\href {\doibase
  10.22323/1.214.0170} {\bibfield  {journal} {\bibinfo  {journal} {PoS}\
  }\textbf {\bibinfo {volume} {LATTICE2014}},\ \bibinfo {pages} {170} (\bibinfo
  {year} {2015})},\ \Eprint {http://arxiv.org/abs/1409.5667} {arXiv:1409.5667
  [hep-lat]} \BibitemShut {NoStop}%
\bibitem [{\citenamefont {Avkhadiev}\ \emph {et~al.}(2019)\citenamefont
  {Avkhadiev}, \citenamefont {Shanahan},\ and\ \citenamefont
  {Young}}]{Avkhadiev:2019niu}%
  \BibitemOpen
  \bibfield  {author} {\bibinfo {author} {\bibfnamefont {A.}~\bibnamefont
  {Avkhadiev}}, \bibinfo {author} {\bibfnamefont {P.~E.}\ \bibnamefont
  {Shanahan}}, \ and\ \bibinfo {author} {\bibfnamefont {R.~D.}\ \bibnamefont
  {Young}},\ }\href@noop {} {\  (\bibinfo {year} {2019})},\ \Eprint
  {http://arxiv.org/abs/1908.04194} {arXiv:1908.04194 [hep-lat]} \BibitemShut
  {NoStop}%
\bibitem [{\citenamefont {Santos}(2017)}]{santos2017ibm}%
  \BibitemOpen
  \bibfield  {author} {\bibinfo {author} {\bibfnamefont {A.~C.}\ \bibnamefont
  {Santos}},\ }\href@noop {} {\bibfield  {journal} {\bibinfo  {journal}
  {Revista Brasileira de Ensino de F{\'\i}sica}\ }\textbf {\bibinfo {volume}
  {39}} (\bibinfo {year} {2017})},\ \Eprint {http://arxiv.org/abs/1610.06980}
  {arXiv:1610.06980 [quant-ph]} \BibitemShut {NoStop}%
\bibitem [{\citenamefont {Lamm}\ \emph
  {et~al.}(2019{\natexlab{b}})\citenamefont {Lamm}, \citenamefont {Lawrence},\
  and\ \citenamefont {Yamauchi}}]{Lamm:2019uyc}%
  \BibitemOpen
  \bibfield  {author} {\bibinfo {author} {\bibfnamefont {H.}~\bibnamefont
  {Lamm}}, \bibinfo {author} {\bibfnamefont {S.}~\bibnamefont {Lawrence}}, \
  and\ \bibinfo {author} {\bibfnamefont {Y.}~\bibnamefont {Yamauchi}} (\bibinfo
  {collaboration} {NuQS}),\ }\href@noop {} {\  (\bibinfo {year}
  {2019}{\natexlab{b}})},\ \Eprint {http://arxiv.org/abs/1908.10439}
  {arXiv:1908.10439 [hep-lat]} \BibitemShut {NoStop}%
\bibitem [{\citenamefont {Tan}\ \emph {et~al.}(2019)\citenamefont {Tan} \emph
  {et~al.}}]{Tan:2019kya}%
  \BibitemOpen
  \bibfield  {author} {\bibinfo {author} {\bibfnamefont {W.~L.}\ \bibnamefont
  {Tan}} \emph {et~al.},\ }\href@noop {} {\  (\bibinfo {year} {2019})},\
  \Eprint {http://arxiv.org/abs/1912.11117} {arXiv:1912.11117 [quant-ph]}
  \BibitemShut {NoStop}%
\end{thebibliography}%

\appendix

\section{OBC Normalization}

Naively, the exponentially-small normalization $\left<\delta_{ij}\right>_{\rho}$ requires $\sim e^{V}$ measurements (in fact, infinitely many for continuous fields). In this appendix we describe how to compute this normalization in polynomial time in the volume. The idea is to interpolate between the OBC and PBC by gradually turning on a term coupling the first and last time-slices ($\Psi_i$ and $\Psi_j$). Chose $f(\alpha)_{ij}$ such that $f(1)_{ij} = \delta_{ij}$ and $f(0)_{ij} = 1$. Defining an interpolating distribution $\rho(\alpha)_{ij} = \rho_{ij} f(\alpha)_{ji}$, we note that the $\rho(1)$ imposes periodic boundary conditions while $\rho(0)$ has the original open boundary conditions. The relative normalization between $\rho(\alpha_i)$ and $\rho(\alpha_j)$ is given by
\begin{equation}
\mathfrak N(\alpha_i,\alpha_j)
= \left< \frac{\rho(\alpha_j)}{\rho(\alpha_i)}\right>_{\rho(\alpha_i)}
= \left< \frac{f(\alpha_j)}{f(\alpha_i)}\right>_{\rho(\alpha_i)}
\text.
\end{equation}
This normalization is multiplicative in the sense that $\mathfrak N(\alpha, \alpha')\mathfrak N(\alpha',\alpha'') = \mathfrak N(\alpha,\alpha'')$. As a result, $\mathfrak N(0,1)\equiv\left<\delta_{ij}\right>_\rho$ can be computed by subdividing it into $N$ terms, $\mathfrak N(0,1)
=\prod_{i=0}^{N-1}
\mathfrak N(\alpha_i,\alpha_{i+1})
$. The performance of this method is sensitive to the choice of interpolating $\alpha$: if the $\alpha_i$ are chosen such that all $\mathfrak N(\alpha_{i},\alpha_{i+1})=e^{-V/N}$ then $\sim N e^{V/N}$ measurements are required. Choosing $N \sim V$, we see that the measurement of $\left<\delta_{ij}\right>_\rho$ can indeed be performed in polynomial time.
\end{document}